\documentclass[runningheads]{llncs}
\usepackage{graphicx}
\usepackage{url}
\usepackage{fontawesome}
\usepackage{setspace}
\usepackage{color}
\usepackage[table,xcdraw]{xcolor}

\begin{document}
\title{Exploring Reproducibility and FAIR Principles in Data Science Using Ecological Niche Modeling as a Case Study}

\titlerunning{Toward Supporting Reproducibility and FAIR Principles in Data Science}

\author{Maria Luiza Mondelli\inst{1}, A. Townsend Peterson\inst{2} \and Luiz M. R. Gadelha Jr.\inst{1}}

\authorrunning{M. L. Mondelli et al.}
\institute{National Laboratory for Scientific Computing, Petr\'opolis - RJ, Brazil\\
\email{\{mluiza,lgadelha\}@lncc.br} 
\and Biodiversity Institute, University of Kansas, Lawrence, Kansas, \\United States of America\\
\email{town@ku.edu}}
\maketitle              

\begin{abstract}
Reproducibility is a fundamental requirement of the scientific process since it enables outcomes to be replicated and verified. Computational scientific experiments can benefit from improved reproducibility for many reasons, including validation of results and reuse by other scientists. However, designing reproducible experiments remains a challenge and hence the need for developing methodologies and tools that can support this process. Here, we propose a conceptual model for reproducibility to specify its main attributes and properties, along with a framework that allows for computational experiments to be findable, accessible, interoperable, and reusable. We present a case study in ecological niche modeling to demonstrate and evaluate the implementation of this framework.
\end{abstract}

\section{Introduction}

Reproducibility is a fundamental aspect of science, and is related to the idea that a scientific process, or experiment, must be able to be reproduced, thus allowing its results to be validated or not \cite{thain2018}. 
If a scientific experiment can be reproduced, it can be reused or extended, leading to the validation of new findings and conclusions.
However, ensuring that an experiment is reproducible is not a trivial task. It involves the need to record detailed documentation and specifications of the whole experimentation process and environment, and hence the need to plan these aspects and how they will be performed.
If these aspects are not strategically thought out before performing the experiment, making it reproducible is time-consuming and unfeasible.
Currently, this challenge has become increasingly important because many experiments are defined by a flow of computational steps. These steps typically perform data processing and prediction activities using machine learning (ML) algorithms to extract knowledge and support decision-making, and are common in data science processes.

In 2016, a survey released by \textit{Nature} \cite{Baker20161500Reproducibility} asked 1500 researchers from different disciplines about reproducibility in research. The study found that 52\% of researchers believed that a crisis exists regarding reproducibility, but most said that they trusted the results of published papers. More than 70\% of the researchers had tried but failed to replicate experiments of other scientists. At the same time, only 20\% said that they had been contacted by another researcher who was trying to replicate their work. Other issues showed that few researchers follow some procedure to allow their experiments to be reproducible; for about 60\% of participants, the most significant challenges include pressure to publish and selective reporting. Other barriers to reproducible research also include the lack of a culture in the research environment where reproducibility is required for all scientific claims \cite{Peng2011ReproducibleScience}. 
In Ecology, the area of the case study of this work, interest in reproducibility is increasing thanks to the use of scripting languages and the need for more open sharing of data \cite{borregaard2016towards}.

Given the importance that reproducibility plays in the scientific process and the relevance of developing methodologies and technologies to support it, here we propose a conceptual model and a framework to formalize its main aspects. 
After introducing the challenge, we describe concepts and present a conceptual model for reproducibility. Next, we propose a framework in which computational experiments can be findable, accessible, interoperable, and reusable (FAIR) and describe a prototype implementation. We evaluate the framework using a case study in ecological niche modeling (ENM). Finally, we explore related work and derive conclusions.
\vspace{-0.2cm}

\section{A Conceptual Model for Reproducibility}\label{cm_reproducibility}

The development of computational experiments that implement common data science steps has become an increasingly common practice in different scientific domains. This increase is thanks to the emergence of new technologies that have supported their execution, especially in cases in which experiments are complex, require high-performance computing, and involve manipulation of large volumes of data. Often, such experiments can be defined as a flow of activities, with execution managed by scientific workflow technologies \cite{Deelman2009}. However each such system has a specificity, which ends up requiring learning time, making it difficult to adopt. Many scientists use scripting languages, such as R and Python, to perform their experiments~\cite{Dey2015LinkingScripts}. 
Both are well-established, open-source languages with broad user communities, and also offer great flexibility in the form of a large number of packages and libraries available, allowing customization of the steps that make up the experiment. Given researches' preference for these languages, recent efforts were applied to support parallel execution of these scripts and collection of provenance data~\cite{Babuji:2019,Pimentel2017}.

Provenance comprises the production history of information or piece of data. By collecting provenance data from an experiment, it is possible to track all steps involved in producing a particular result. 
Scientists can benefit from provenance information to verify the sequence of computational steps performed to produce an output, to validate whether this sequence was executed correctly and used the appropriate parameters and inputs, and to compare different executions of the same experiment. Information about versions of libraries or packages, their dependencies, and aspects of the environment can also help in configuration of an environment more similar to the original when reproducing an experiment.

\vspace{-0.4cm}
\begin{figure}[ht]
\centering
\includegraphics[width=0.8\textwidth]{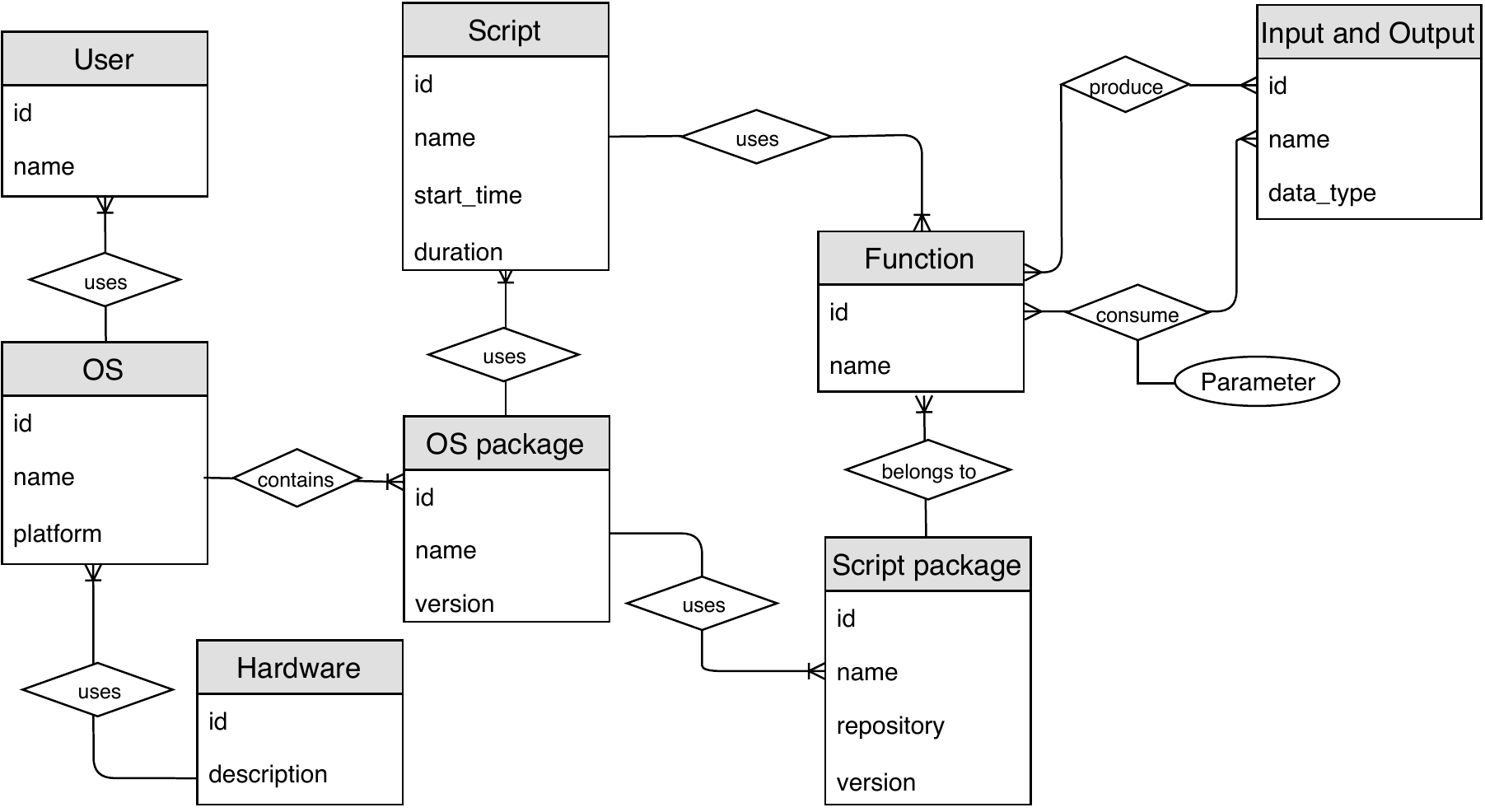}
\caption{Entity-Relationship diagram representing the aspects involved in the process of reproducibility of computational experiments.} \label{fig:model}
\end{figure}
\vspace{-0.4cm}

In the process of designing, implementing, and executing an experiment, several aspects must be taken into account to allow it to be reproducible. In this work we propose a conceptual model that maps aspects that we consider essential for reproducibility, especially for cases in which experiments are modeled as workflows through scripting languages. This model is presented in the form of an entity-relationship diagram, as shown in Fig.~\ref{fig:model}, with aspects we identified as essential mapped as entities. The main idea behind the model and the relationships between its entities is as follows: a \textit{user} specifies and runs his/her
experiment from an \textit{operating system} (OS) using a specific \textit{hardware}. This OS has some packages installed (\textit{OS package}), and among them, we can consider the scripting language package that is used to specify the experiment, such as R or Python. The experiment is defined through a \textit{script}, from an existing package in the OS, and can contain calls to user-defined functions or functions (\textit{function}) that are part of a specific language package or module (\textit{script package}). These specific language modules need to be installed in the OS. The functions comprise the activities of the experiment that consume and produce \textit{input} and \textit{output}, respectively. 
Parameters can also be used as input to functions, and therefore constitute an attribute of the \textit{consume} relation.

\begin{figure}
\includegraphics[width=\textwidth]{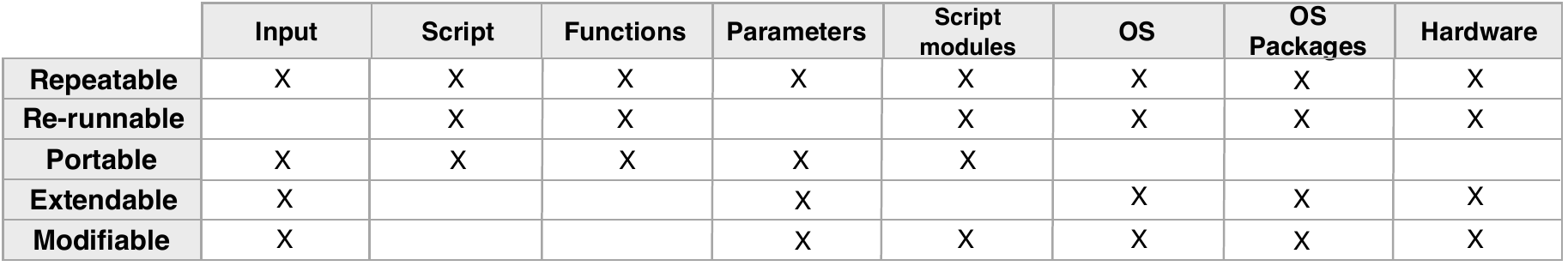}
\caption{Levels of reproducibility and the respective entities involved (adapted from~\cite{Freire2018}).} \label{fig:layers}
\end{figure}

The record of the aspects described through the entities of the proposed model can support the process of reproducibility on different levels, as addressed in~\cite{Freire2018}. We adapted the level definitions to cover the entities that we included in the model, as shown in Fig.~\ref{fig:layers} and described as follows:

{\bf Repeatable}: executions of an experiment in the same computing environment, using the same code and data set, produce the same results or results consistent with the original finding. Thus, we must preserve all entities described in the proposed model.
    
{\bf Re-runnable}: the results of an experiment remain consistent even with variations in the input data or in the parameter settings used. For this reason, the implementation that concerns the script and function entities should not be modified, as well as the hardware, module packages, operating system, and its packages.
    
{\bf Portable}: the experiment does not depend on the computing environment in which it was originally executed. Aspects such as versioning and availability of libraries and dependencies should be considered at this level. We must then preserve the input data, script, its functions, parameters, and module packages. 
    
{\bf Extendable}: an experiment can be reused for other purposes. This reuse includes the possibility of integrating the original experiment into a new or existing experiment. For this aspect, one must have access to the script and its functions. We can preserve the operating system, its packages, and module packages.
   
{\bf Modifiable}: changes can be made in the implementation of the original experiment for reuse purposes. As with the previous level, it is necessary to have access to the scripts and functions of the original experiment.

We emphasize that by saying that we should preserve some entity, we mean that it should not be changed to ensure reproducibility at a certain level. We can observe that the level of repeatability for which we are aiming is the narrowest, and guarantees what we can call exact reproducibility. However, it is possible to combine aspects of more than one level depending on the desired degree of reproducibility.

\vspace{-0.2cm}
\vspace{-0.8cm}

\section{A Framework for FAIR Computational Experiments}\label{framework}
The FAIR principles \cite{wilkinson2016} emerged from the efforts of a community including individuals from academia, industry, funding agencies, and scholarly publishers, to establish guidelines that allow for the findability, accessibility, interoperability, and reuse of scientific data or digital assets in general. Each of these four principles contains a set of requirements on how data, metadata, and infrastructure must be managed, allowing machines to retrieve them automatically, or at least with minimal human intervention. From a reproducibility point of view, the existence of such a guideline is beneficial. Although the principles do not suggest any specific technology to be used, they provide a means by which to solve aspects related to, for example, obtaining the input data for an experiment.
However, as seen in the previous section, the reproducibility of an experiment is not limited to aspects related to its data. If we want to support the process of making an experiment reproducible, we can apply the FAIR principles to the experiment in its entirety. Therefore, each of the entities described in the conceptual model (Fig.~\ref{fig:model}) must have enough information to enable the user to reproduce the experiment at the desired level (Fig.~\ref{fig:layers}).

\begin{figure}
\includegraphics[width=\textwidth]{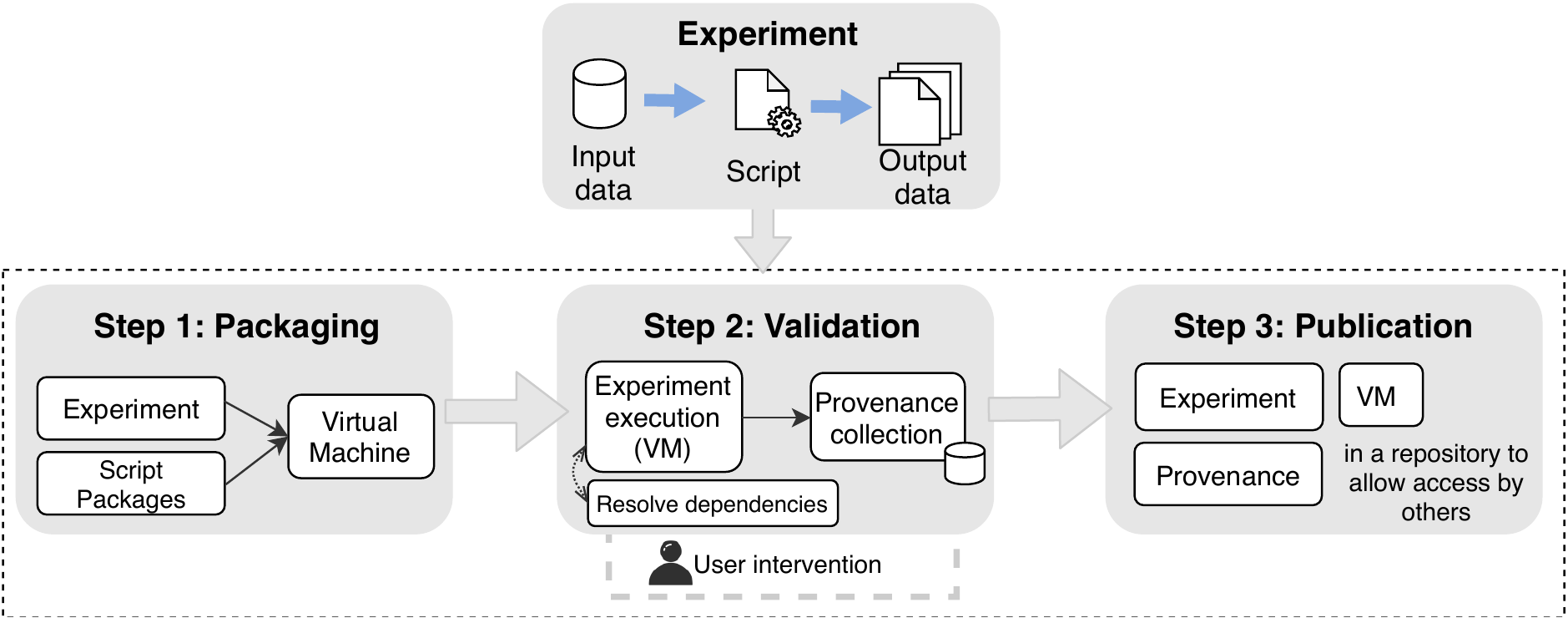}
\caption{Framework for FAIR computational experiments and its components.} \label{fig:framework}
\end{figure}

Considering the scenario in which the user has implemented an experiment through a scripting language, we propose a framework (Fig. \ref{fig:framework}) to make this experiment reproducible. The main idea behind the framework is that, by following the proposed steps, the user will be able to share the experiment with enough information that meets, even minimally, the FAIR principles.

\vspace{-0.4cm}
\vspace{-0.4cm}

Then, starting from an experiment that consumes input data, processes those data through the execution of a script like R or Python, and generates output data, the first step is to allow the encapsulation (or packaging) of that experiment. This packaging consists of importing the experiment and the packages or libraries used into a virtual machine (VM) using a standard operating system. 
It is important that the scripting language includes some dependency management system. With such a system, it is possible to identify which packages need to be included in the VM, avoiding installation of unused dependencies.

The second step is to access the VM created in the previous step and re-execute the experiment in this new environment. This step requires user interaction, which is responsible for validating the generated results. 
The user is responsible for resolving potential issues that may occur, such as installing the applications required to run the experiment and not previously installed in the VM. All modifications made to the VM, especially as regards installation of the new applications, must be recorded. This record will allow to update the default VM specifications to include modifications. Thus, a user intending to reproduce the experiment will be able to rebuild the virtual environment in which it has been validated, without the need to install the required applications manually. During the experiment re-execution, it is also interesting to provide in the framework a way to collect provenance data and store, for example, into a database that can be subsequently queried for verification or validation purposes. Both R and Python have packages that can support this process.

Finally, the third step aims to gather the provenance database in the previous step, the experiment, and the VM specifications for publication. Publishing can be done in a repository that allows the sharing of research results. The choice of which repository is ideal for such sharing may vary depending on the application area of the experiment. However, initiatives and platforms such as FAIRsharing\footnote{https://fairsharing.org/} and Repository Finder\footnote{https://repositoryfinder.datacite.org/} have made available ways to support this process through recommendations and tools for searching repositories.

To exemplify the framework implementation for experiments defined with R,
we encapsulated each step in R functions using a set of available R packages that meet the requirements to allow reproducibility and the FAIR principles of the experiment. These functions and a guide for their use are available at GitHub\footnote{https://github.com/mmondelli/reproduceR}. We start from the scenario in which the workflow has already been implemented and executed by the user. The following describes each of the steps and tools used:

{\bf Step 1}. For the packaging of the experiment, we use a package that manages the dependencies of the experiment called \textit{packrat}. \textit{Packrat} stores, in the experiment directory, the package installation files in the versions that were used in the script. Also, \textit{packrat} has a bundle function, which compresses these packages and the script for later restoration. We then use \textit{Vagrant}, a tool for building and managing VM environments, to create and start an Ubuntu VM containing the R installation for running the experiment. This step is achieved from the experiment directory, which is shared by default with the created VM, so that one can access the bundled experiment when connecting to the machine.

{\bf Step 2}. In the validation step, we use \textit{packrat} to unpack the experiment and install the packages. 
The installation of specific applications required by the experiment must be resolved by the user manually. We then provide a function that re-executes the experiment using the \textit{rdt} package, responsible for collecting the provenance. In this same function, we create a relational database (according to Fig. \ref{fig:model}) to hold the provenance information collected.

{\bf Step 3}. Finally, we put together the relational database containing the provenance, the bundled experiment generated by \textit{packrat}, and the VM specification, which includes the installation of the new applications, and we compress these files for publication. We then use the \textit{zen4R} package to publish on the Zenodo platform. Once this step is done, users can access the platform to add the necessary information and details about the publication.
\vspace{-0.3cm}

\section{Case Study in ENM and Evaluation}\label{Model-R}

ENM is a technique that, based on data describing environmental factors, seeks to identify environmental requirements and predict the geographic distribution of a species. The environmental factors are generally abiotic, including aspects such as temperature, elevation, and humidity, but can also be biotic, describing the interactions that species establish with one another. Together they shape the environmental and geographic distributions of the species. The technique has been used widely in recent years, and has been applied in studies in conservation, ecology, and evolution. 

The ENM process relies on the use of statistical and ML tools that allow  analysis of environmental variables in relation to the geographic coordinates that represent the points of occurrence of a species. 
The end result is identification of a set of conditions associated with the occurrence of the species in question.
Model-R \cite{sanchez2017model} is a workflow, implemented in R, that automates some of the common steps for performing ENM. 
The following is an overview of each of these steps and the respective activities implemented by Model-R (Fig. \ref{fig:workflow}):

\begin{figure}
\includegraphics[width=\textwidth]{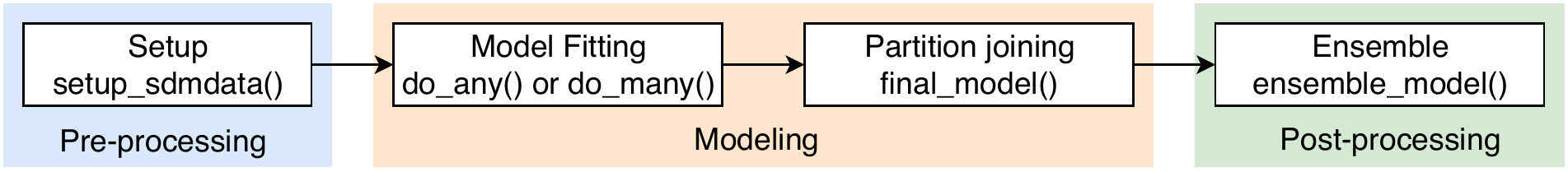}
\caption{Model-R workflow activities.} \label{fig:workflow}
\end{figure}
\vspace{-0.5cm}

{\bf Pre-processing}: comprises the data acquisition stage, which includes assembly of environmental layers and occurrence points, used as input to model algorithms. These data are usually obtained through known databases and require a cleaning step to eliminate uncertainties, duplications, and inconsistencies. In Model-R, \textit{Setup} implements the data cleaning and partition procedures.

{\bf Modeling}: consists of application of algorithms that generate the ecological niche models, based on the data obtained in the previous stage. In general, these algorithms use associations between the points of occurrence of the species and the environmental layers to predict the potential for occurrence of the species in other areas. This step is often achieved by using ML algorithms such as RandomForest and Maxent. This step is implemented in Model-R in two steps: (i) \textit{Model Fitting}, which creates the ENM for a particular instance; and (ii) \textit{Partition Joining}, which joins multiple models into a single result per species.

{\bf Post-processing}: consists of the evaluation of the model generated in the previous step, to establish whether it is adequate and robust. Models can be evaluated using statistical methods, based on occurrence data not used in the preceding step. The \textit{Ensemble} activity in Model-R joins the models for each algorithm into a final model, also known as a consensus model, which can then be analyzed and evaluated by the user.

To evaluate the implementation of the framework, we used an example of the Model-R workflow, available in the documentation\footnote{https://github.com/Model-R/modelr\_pkg} for modeling {\em Abarema langsdorffii} (Benth.) Barneby \& J.W.Grimes, a plant species for which data are available on the Model-R package. We executed the experiment on a machine locally, and followed the steps of the proposed framework for creating the VM. 
We highlight that, in the second step of the framework, we manually resolved one of the dependencies: the installation of the GDAL application in the VM.
Next, we verified if it is possible to reproduce the experiment considering two different scenarios: (i) the user wants to reconstruct and reproduce the experiment locally, installing the packages used in the script, and not necessarily using the same versions; and (ii) the user wants to reproduce the experiment from a VM created by the framework. 

\begin{table}
\caption{Results of the framework evaluation.}\label{results}
\centering
\begin{tabular}{c|c|c|c|}
\cline{2-4}
 & \textbf{\begin{tabular}[c]{@{}c@{}}Machine 1\end{tabular}} & \textbf{Machine 2} & \textbf{Machine 3} \\ \hline
\multicolumn{1}{|c|}{\textbf{Config./Scenario}} & \begin{tabular}[c]{@{}c@{}}Ubuntu 18.04 LTS\\ R 3.6.0\end{tabular} & \begin{tabular}[c]{@{}c@{}}CentOS Linux 7\\ R 3.4.4\end{tabular} & \begin{tabular}[c]{@{}c@{}}Windows\\ R 3.4.4\end{tabular} \\ \hline
\multicolumn{1}{|c|}{\begin{tabular}[c]{@{}c@{}} Local \footnotesize{(no framework)}\end{tabular}} & \faTimes & \faTimes & \faTimes \\ \hline
\multicolumn{1}{|c|}{\begin{tabular}[c]{@{}c@{}}VM \footnotesize{(with framework)}\end{tabular}} & \faCheck & \faCheck & \faCheck \\ \hline
\end{tabular}

\end{table}

For this step, we used other machines with distinct hardware configurations. We took an intermediate file (sdmdata.txt), generated during the execution of the \textit{Setup} step, as an example to verify if its content changed according to the machine used and the scenario in which the file was generated. This file contains clean data that are consumed by subsequent activity. The \textit{Setup} activity has a random process and a seed is used as a parameter to ensure reproducibility. However, we were able to verify that only in the second scenario, with the use of the VM, 
the contents of the file remained the same when compared to the machine that originated the experiment (Table \ref{results}). The X symbol represents the executions that produced different results when compared to the machine that originated the experiment.
The implementation of the framework and the execution of the experiment using the VM was robust enough to guarantee that the results were the same as the original, regardless of the machine used.

\section{Related Work}\label{related}

A recent survey \cite{thain2018} raises the reasons why reproducibility of computational experiments is needed and the technical barriers and challenges that exist to achieve it. These barriers include issues ranging from managing and recording information from the execution environment to aspects related to the data that are consumed and generated by those experiments. Some efforts and initiatives have attempted to address these issues, especially as journals are increasingly encouraging submissions of reproducible computational research \cite{stodden2013toward}.

Among these efforts, we highlight \textit{encapsulator} \cite{encapsulator}, a toolbox that relies on provenance data to produce an environment in which computational experiments can be reproduced. 
\textit{ReproZip} \cite{reprozip} is a packaging tool that also uses provenance and focuses primarily on identifying the dependencies required to run an experiment. 
\textit{WholeTale} \cite{wholetale} is a computational environment with features for data collection, identity management, data publication, and interfaces to analytical tools. 
These tools, as well as the framework proposed in this work, rely on VMs to allow reproduction of the experiment by others. However, we emphasize that in this work, we do not seek to implement a specific tool to guarantee reproducibility. Instead, we propose a framework that can benefit from existing tools to achieve different levels of reproducibility. We extend the proposal of \textit{encapsulator} and \textit{ReproZip} by indicating the need to publish the aspects related to the experiment so that it can be uniquely identified, accessed, and retrieved later. This work extends our previous work on provenance management \cite{Mondelli2018a}. The implementation of our framework provides provenance data, VM specifications, experiment script, and package versions used as a result. The user intending to run the experiment is not limited to using the VM, but can manually retrieve and reuse the experiment, or query the provenance database.

Addressing the FAIR principles, Madduri et al. \cite{madduri2019} presented a set of tools that can be used to support the implementation of computational experiments, and ensure the aspects related to each of the principles. Madduri et al. \cite{madduri2019} used a case study in biomedicine in which data and computation are often distributed. As in this paper, the need to apply FAIR principles in the experiment as a whole is discussed. However, in this work, we seek to present a more general and higher-level view of aspects related to computational experiments that can be registered and the steps that can be followed to achieve reproducibility. In this way, the conceptual model and the framework can be more easily adapted to the specific needs of each experiment.

\section{Conclusions}\label{conclusions}

In this paper, we discussed aspects of the reproducibility of computational experiments and the importance of allowing the application of FAIR principles to the experiment in its entirety. We indicate, through a conceptual model, the essential aspects of the experiments that must be recorded and how they relate to different levels of reproducibility. Also, we propose a generic framework for FAIR computational experiments and demonstrate its implementation through an ENM case study.
As future work, we intend to evaluate and quantify which of the FAIR metrics 
are met with the implementation of the proposed framework. Future work may also include a mapping between the metrics and the different levels of reproducibility, and evaluation of what level of reproducibility is achieved according to the metrics obtained.
We also intend to study the uncertainty quantification of the different results generated by the ML algorithms of the ENM experiment. \\
\textbf{Acknowledgments}. This work was supported by CNPq, CAPES, and FAPERJ. We thank Marinez Ferreira, Andrea S\'anchez-Tapia and Sara Mortara, from the Botanic Garden of Rio de Janeiro, for their contributions.

\bibliographystyle{splncs04}
\bibliography{ref}

\end{document}